\documentclass[reprint,superscriptaddress,floatfix,raggedbottom,aps,prl]{revtex4-2}
\usepackage{amsmath}
\usepackage{amssymb}
\usepackage{graphicx}
\usepackage[hidelinks]{hyperref}

\begin{document}

\title{\texorpdfstring{From Atomistic Impurity Landscapes to Two-Level-System Distributions and Microwave Loss: Hydrogen-Induced Losses in Nb$_2$O$_{5-\delta}$ and Ta$_2$O$_{5-\delta}$}{From Atomistic Impurity Landscapes to Two-Level-System Distributions and Microwave Loss: Hydrogen-Induced Losses in Nb2O(5-delta) and Ta2O(5-delta)}}

\author{Crist\'obal M\'endez}
\affiliation{School of Applied and Engineering Physics, Cornell University, Ithaca, New York 14853}

\author{Sarvesh Chaudhari}
\author{Tom\'as A. Arias}
\affiliation{Department of Physics, Cornell University, Ithaca, New York 14853}

\date{October 15, 2026}

\begin{abstract}
We introduce a first-principles-guided framework for predicting statistical TLS distributions from mobile impurities in disordered materials. Machine-learned interatomic potentials, with DFT and minimum-action-path validation, connect impurity landscapes to tunnel-splitting and asymmetry distributions, TLS densities, and dielectric loss. Applied to amorphous Nb$_2$O$_{5-\delta}$ and Ta$_2$O$_{5-\delta}$, the framework implicates interstitial hydrogen as a source of microwave loss. Loss tangents match Nb-pentoxide measurements at inferred H concentrations, while weaker H uptake in Ta reproduces the Nb/Ta TLS contrast.

\end{abstract}

\maketitle

Tunneling two-level systems (TLSs), in which a condensed-matter system tunnels between two nearly degenerate configurations, were introduced to explain anomalous low-temperature properties of glasses and remain the canonical description of low-energy excitations in disordered solids~\cite{Anderson1972,Phillips1972,phillips1987two}. The microscopic identity of the tunneling centers, however, has remained elusive for decades. Atomistic simulations have identified candidate tunneling motifs in amorphous alumina and statistical quantum-TLS ensembles in model glasses~\cite{HeuerSilbey1993,PerezPaz2014,Khomenko2020,Ciarella2023}. Related energy-landscape approaches have connected thermally activated double-well defects to mechanical loss in silica, Zr-doped tantala, and amorphous Si~\cite{DamartRodney2018,Jiang2021,Girard2026}. At microwave frequencies, the outstanding challenge is to connect realistic atomistic energy landscapes to distributions of quantum tunnel splittings and experimentally observable dielectric loss in specific materials.

This challenge is particularly pressing in superconducting qubits and superconducting radio-frequency (SRF) cavities, where TLS in surface and interface dielectrics contribute to microwave dissipation and noise~\cite{Mueller2019,RomanenkoSchuster2017}. Niobium and tantalum are central superconducting materials in these technologies, bringing their native oxides into focus as possible hosts of the relevant tunneling centers~\cite{Verjauw2021,Chang2025}. We therefore turn to amorphous Nb$_2$O$_{5-\delta}$ and Ta$_2$O$_{5-\delta}$, and to the experimental constraints that any microscopic assignment must satisfy.

Experiments on Nb and Ta oxides provide several design rules for a microscopic TLS model. First, TLS loss is strongly correlated with structural disorder: amorphous oxides exhibit larger loss than more ordered oxides~\cite{oh2024structure}, motivating amorphous structural ensembles. Second, the loss appears to be associated specifically with the pentoxide rather than with all niobium oxides equally. Specifically, microwave loss tracks Nb$_2$O$_5$ regrowth in native Nb oxides~\cite{Verjauw2021}, oxygen-vacancy-associated TLSs have been implicated in Nb pentoxide~\cite{Bafia2024}, and isolated NbO$_2$ shows no detectable TLS loss in bulk-powder measurements~\cite{ganesan2026two}; these observations motivate pentoxide hosts with oxygen-deficient configurations. Third, TLS loss differs between the two pentoxides, with Nb$_2$O$_5$ producing greater resonator loss than Ta$_2$O$_5$ in otherwise similar devices~\cite{goronzy2026comparison}, motivating a direct Nb/Ta comparison. Together, these observations define the material space and design levers addressed here, but they do not uniquely identify the microscopic TLS. Several defect classes remain plausible~\cite{Mueller2019}, and recent studies have focused mainly on electronic defects, including local moments in substoichiometric pentoxide~\cite{pritchard2025suppressed} and dangling-bond--like centers~\cite{wang2025superconducting}.

Impurity tunneling centers provide another plausible class of TLSs. For example, near-surface hydrogen has been measured at Nb oxide/Nb interfaces~\cite{romanenko2011elastic} and hydride formation has been observed in Nb qubit films~\cite{sung2026formation}. Moreover, owing to its small mass and the polarity of the bonds it forms, hydrogen can tunnel between nearby configurations while coupling strongly to electromagnetic fields.

The core difficulty in studying such tunneling centers from first principles is that microwave-active TLSs occupy an exceptionally narrow $\mu$eV energy window. Their tunnel splittings depend exponentially on barrier shape, hop distance, and effective mass, while microwave activity also requires asymmetries on the same minute scale. Present-day \textit{ab initio} calculations therefore cannot reliably predict the parameters of an individual disordered tunneling TLS: small errors in the energy landscape can produce orders-of-magnitude errors in its tunnel splitting, and $\mu$eV asymmetries cannot be resolved directly. The rarity of such TLSs further makes exhaustive first-principles sampling infeasible. Material-specific studies have consequently focused mainly on assessing selected motifs rather than predicting statistically representative microwave-active ensembles~\cite{DuBois2013,Holder2013,Gordon2014}.

We instead adopt a statistical, multifidelity approach. We use MLIP-accelerated sampling to map the metastable environments and transition landscape of a chosen mobile impurity, targeted DFT calculations to anchor representative minima and barriers, minimum-action-path and Schr\"odinger calculations to benchmark a reduced tunneling model, and statistical coarse-graining to convert the resulting ensemble into frequency-resolved TLS densities and concentration-dependent loss tangents. The underlying ansatz, tested against collective observables such as measured loss tangents, is that although theory cannot reliably determine which specific pairs fall within the microwave window, it can capture the broader asymmetry and tunnel-splitting distributions well enough that errors mainly reshuffle microwave-active pairs rather than substantially alter their ensemble density. Applied here to amorphous Nb and Ta pentoxides, our framework establishes that hydrogen-related TLSs are consistent with key experimental trends for the observed microwave losses in the $\sim0.1$--$10$\,GHz range.

\begin{figure}[!htbp]
   \centering
   \includegraphics[width=1.0\columnwidth]{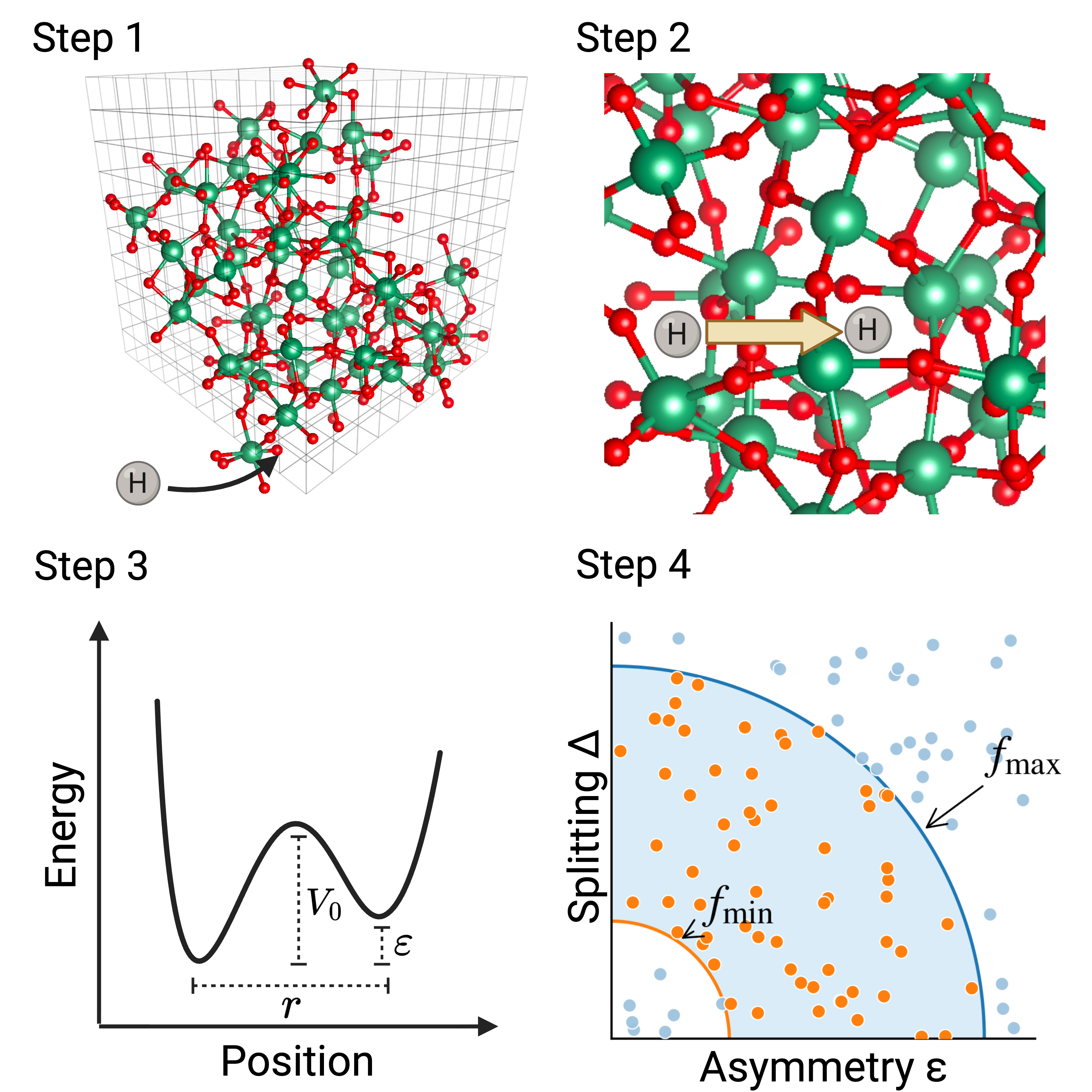}
\caption{Workflow for identifying tunneling TLS in disordered materials. (1) Sample, relax, and cluster trial defects into metastable minima. (2) Select nearby, nearly symmetric pairs. (3) Compute tunneling paths to obtain barrier, asymmetry, and splitting. (4) Map the ensemble into reduced TLS-parameter space and integrate over the microwave band to estimate spectral density and intrinsic loss. Step 4 is schematic, not calculated data.
}
   \label{fig:tls_identification_workflow}
\end{figure}

\noindent \emph{A transferable statistical workflow for mobile-impurity tunneling TLS.---} Figure~\ref{fig:tls_identification_workflow} outlines our workflow for mobile impurity tunneling. We (1) sample and relax trial configurations of inserted impurities to catalogue distinct metastable local minima, (2) select nearby, low-asymmetry minima as candidate tunneling pairs, (3) determine the tunneling profile connecting each pair and extract the corresponding TLS parameters, and (4) map the resulting ensemble into reduced TLS-parameter space to estimate the microwave-active TLS density and intrinsic loss tangent. MLIP-accelerated sampling makes the first three steps tractable over large disordered search spaces, targeted DFT calculations anchor representative local environments and barriers, and our final statistical step converts atomistic tunneling-pair ensembles into experimentally relevant observables.

\noindent \emph{Defect type.---}
To proceed with this general approach, we must first identify the likely class of defects responsible for the observed losses. For Nb devices, we begin by asking which interstitial species is most plausible as a microwave-regime tunneling center. As a simple initial screening, we consider hydrogen, interstitial oxygen, and nitrogen tunneling between their corresponding interstitial sites.

\begin{figure}[!htbp]
  \centering
  \includegraphics[width=0.92\columnwidth]{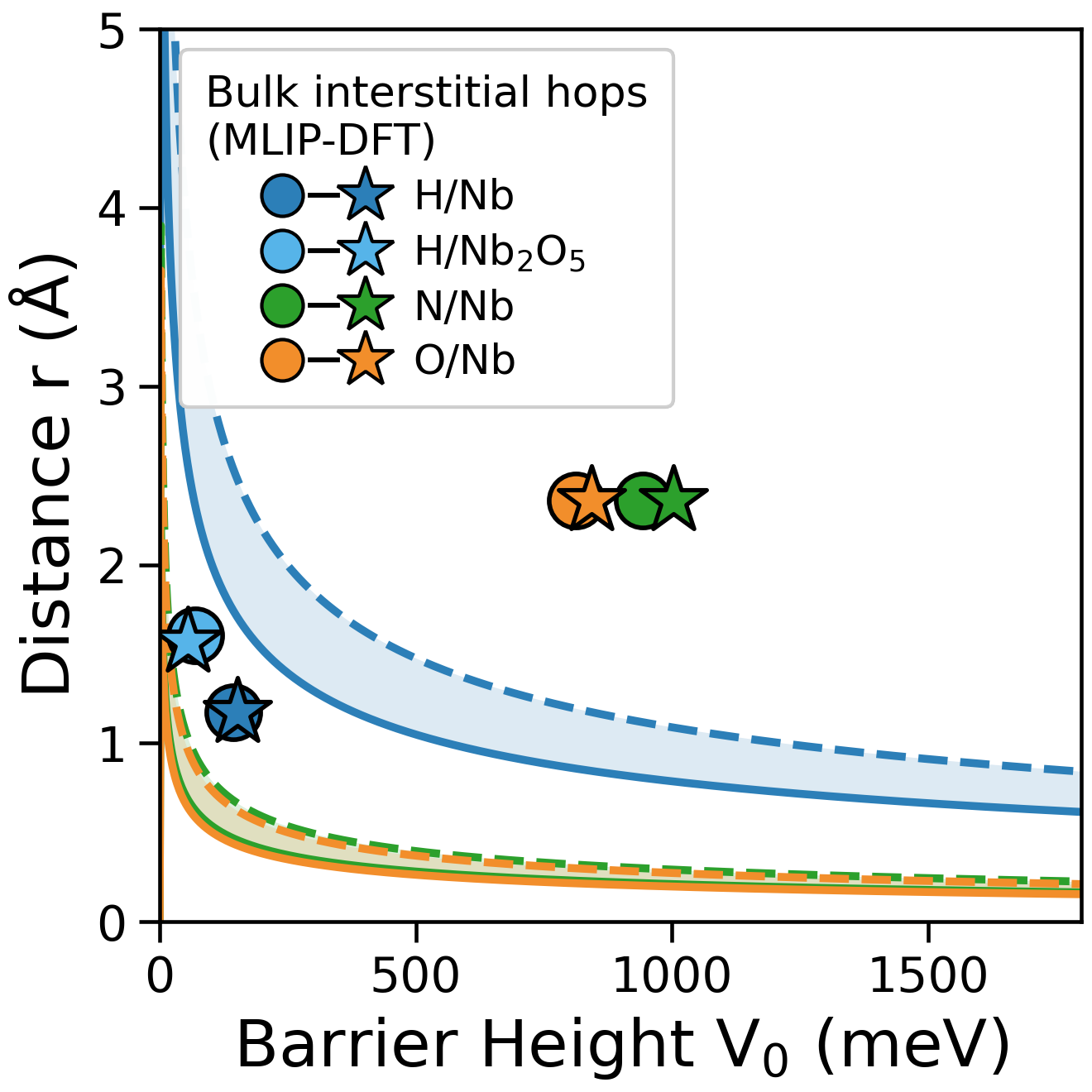}
  \caption{Barrier height versus hop distance for representative H, N, and O interstitial hops in bulk Nb and H interstitial hops in bulk Nb$_2$O$_5$. Circles and stars denote MLIP and DFT values, respectively. Solid and dashed curves mark quartic-WKB isofrequency contours for $10$ and $0.1$~GHz, respectively, using the tunneling mass of each species, and the shaded regions highlight the corresponding $0.1$--$10$~GHz windows. The representative hydrogen points, including those in Nb$_2$O$_5$, lie on the high-frequency side of the hydrogen window (smaller $r$ at fixed $V_0$), but much closer to the microwave-active regime than the oxygen and nitrogen points.}
  \label{fig:bulk_nb_interstitial_benchmark}
\end{figure}

Figure~\ref{fig:bulk_nb_interstitial_benchmark} shows the hop distances $r$ and barrier heights $V_0$ for representative interstitial tunneling events in bulk Nb for H, N, and O, computed with both DFT and the FairChem UMA MLIP~\cite{wood2025uma,fairchem_github}. Because this initial screening uses bulk Nb, the asymmetry is zero, $\varepsilon=0$. The agreement between the DFT and MLIP points supports the use of the MLIP to locate these defects in the distance--barrier phase space. The figure also overlays quartic-WKB isofrequency contours corresponding to tunnel splittings of 0.1 and 10~GHz for each species, using the appropriate tunneling mass. Because nitrogen and oxygen have similar masses, their contours nearly overlap. Our results show that oxygen and nitrogen are implausible candidates for TLS in the 0.1--10~GHz microwave regime: their larger hop distances, higher barriers, and larger atomic masses place them far outside the microwave-active window, unless their hop distances were reduced to the sub-\AA{} scale or their barriers were lowered far below 0.1~eV. Hydrogen in Nb, by contrast, lies much closer to the relevant frequency window. Moreover, the results for H tunneling in bulk Nb$_2$O$_5$ shown in Figure~\ref{fig:bulk_nb_interstitial_benchmark} (corresponding to the lowest-barrier path connecting degenerate lowest-energy interstitial sites) likewise lie reasonably close to the H 0.1--10~GHz band. This proximity makes it plausible that local disorder in amorphous oxides could shift some H defects into the microwave-active window. We therefore focus the remainder of this study on hydrogen interstitial defects as plausible TLS candidates in the relevant frequency range.

Having identified the relevant defect class, we now describe the successive steps used to construct the TLS ensemble and analyze its statistical properties to predict observable microwave losses, as illustrated in Figure~\ref{fig:tls_identification_workflow}.

\noindent\emph{Step 1: Local Minima.---} We must first catalogue a representative set of local minima that can potentially form tunneling pairs. For interstitial defects in niobium- and tantalum-oxide hosts, we begin with representative structures and then add hydrogen interstitials on a uniform sampling grid, allowing the system to fully relax in each case. We choose the grid spacing $h_{\mathrm{grid}}$ to be sufficiently fine to resolve distinct local basins after relaxation. Finally, to avoid multiple counting of minima, we cluster the relaxed structures into numerically identical minima and select a single representative of each cluster.

Here, we focus on TLS in oxygen-deficient amorphous niobium and tantalum pentoxides, M$_2$O$_{5-\delta}$ ($M=\mathrm{Nb},\mathrm{Ta}$). For our calculations, we use a uniform H-seeding grid with spacing $h_{\mathrm{grid}}=1$~\AA{} across 49 amorphous cells per oxide from Pritchard \textit{et al.}~\cite{pritchard2025suppressed}, spanning seven oxygen-deficiency levels. This yields 7,137 distinct H minima in Nb$_2$O$_{5-\delta}$ and 7,825 in Ta$_2$O$_{5-\delta}$. The H interstitial formation-energy distributions remain systematically lower in Nb$_2$O$_{5-\delta}$ than in Ta$_2$O$_{5-\delta}$ at every oxygen-deficiency level $\delta$ (see End Matter). This stronger thermodynamic tendency for hydrogen incorporation in Nb pentoxide further motivates considering lower H incorporation in Ta when comparing loss below.

\noindent\emph{Step 2: Tunneling Pairs.---} We next identify local-minimum pairs that can form double wells, requiring proximity in both space and energy. We therefore screen pairs using a maximum site-to-site separation $r_{\mathrm{cut}}$ and a maximum asymmetry magnitude $\varepsilon_{\mathrm{cut}}$.

For our application, we define candidate double wells as distinct hydrogen minima with relaxed site-to-site separation $r\le r_{\mathrm{cut}}=1.65$~\AA\ and asymmetry magnitude $|\varepsilon|\le \varepsilon_{\mathrm{cut}}=10$~meV. These thresholds isolate nearby, nearly degenerate hydrogen configurations that can plausibly act as tunneling pairs while retaining a sufficiently large disorder-generated ensemble for statistical analysis.

\noindent\emph{Step 3: Tunneling Splittings.---} Having identified candidate tunneling pairs, we next assign a tunnel splitting $\Delta$ to each double well. In general, this requires determining the effective tunneling coordinate connecting the two minima, reconstructing the potential along that coordinate, and solving or approximating the resulting one-dimensional quantum problem. The highest-fidelity version of this procedure is to compute the fixed-energy optimal tunneling path (OTP), which identifies the minimum-action path (MAP) controlling the semiclassical tunneling amplitude~\cite{Asgeirsson2018}. This path better represents the quantum-mechanical tunneling trajectory than the adiabatic minimum-energy path (MEP) alone. The optimized OTP segment can then be joined to harmonic endpoint tails obtained from endpoint force constants, and the tunnel splitting $\Delta$ can be obtained by solving the full one-dimensional Schr\"odinger equation on the reconstructed potential.

For our amorphous oxide ensemble, applying this full OTP-based workflow to every candidate pair is computationally prohibitive with our current methods. Moreover, our goal is the statistical distribution of TLS parameters rather than isolated high-accuracy splittings for a few selected defects. We therefore use the full OTP/Schr\"odinger procedure as a benchmark and validation tool and a reduced model for ensemble screening. Following prior first-principles TLS studies in alumina and bulk Nb~\cite{Holder2013,Abogoda2025}, we approximate each tunneling profile by the adiabatic minimum-energy path (MEP) obtained using the nudged elastic band (NEB) method~\cite{jonsson1998nudged}. To retain the physics of the minimum-action path in this ensemble-scale approximation, we supplement the NEB treatment with an effective mass calibrated to reproduce the tunnel splitting obtained from a full OTP calculation for a representative tunneling center, as described below.

\begin{figure}[!htbp]
\centering
\includegraphics[width=1.0\columnwidth]{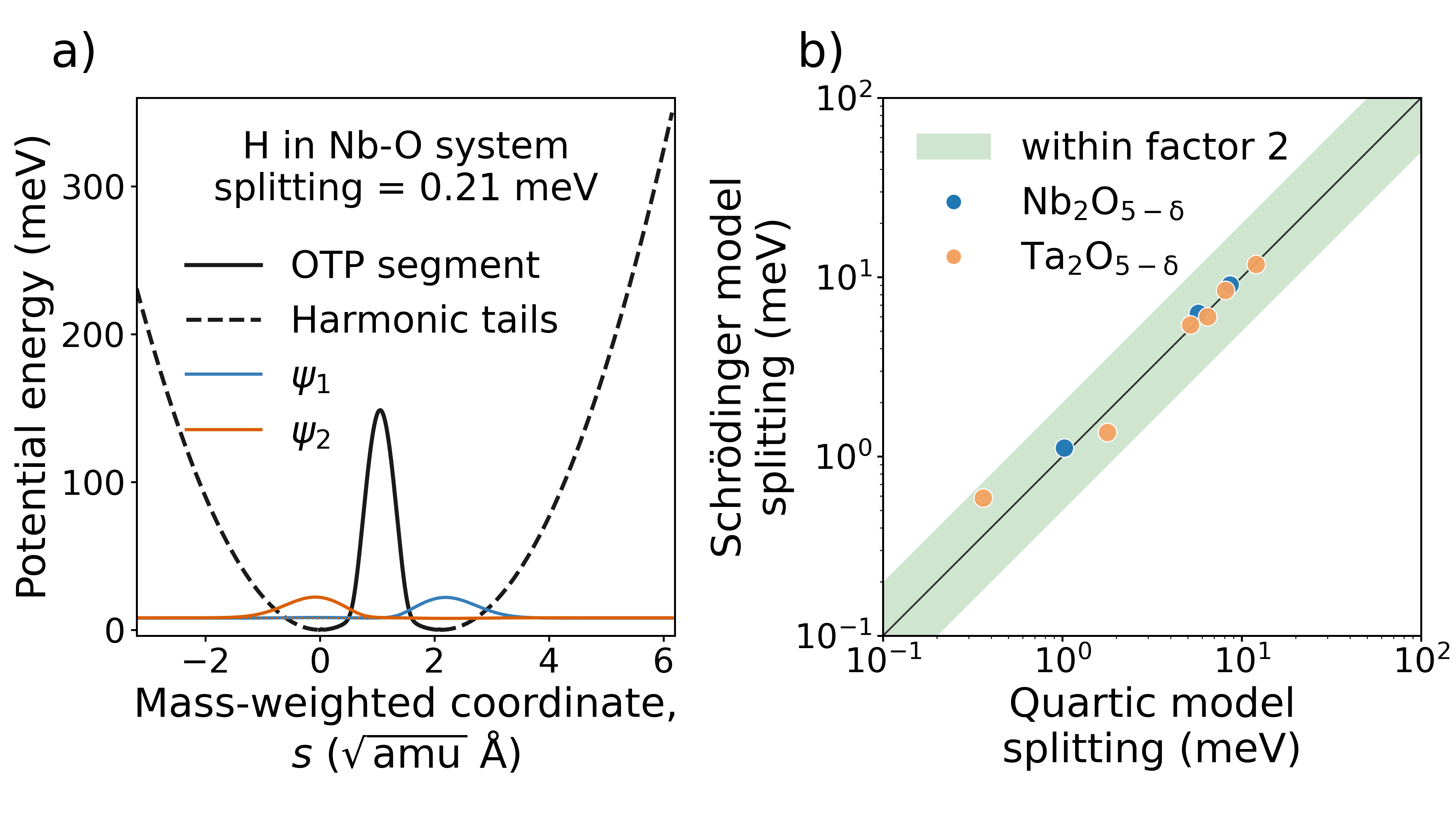}
\caption{Tunneling-path and NEB-WKB validation benchmarks. (a) One-dimensional Schr\"odinger benchmark for oxygen-trapped interstitial hydrogen in bcc Nb. The optimized tunneling path (OTP) segment is joined to harmonic endpoint tails obtained from endpoint force constants, producing the reconstructed double-well potential in the mass-weighted coordinate $s$; solving the one-dimensional Schr\"odinger equation gives a $0.21$~meV splitting, consistent with the $0.19\pm0.02$~meV experimental O--H tunneling scale~\cite{wipf1981neutron,wipf1984h}. (b) OTP-Schr\"odinger versus NEB-WKB comparison for representative Nb$_2$O$_{5-\delta}$ and Ta$_2$O$_{5-\delta}$ hydrogen-tunneling paths. The diagonal gives exact agreement; green band marks agreement within a factor of 2.}
\label{fig:tunneling_validation_benchmarks}
\end{figure}

For a quantitative comparison, we use the experimentally established O--H tunneling defect in bcc Nb, where H is trapped by a local Nb--O environment and the tunneling scale is directly measured. Inelastic neutron scattering on dilute NbO$_{0.013}$H$_{0.016}$ found an O--H tunneling matrix element of $0.19\pm0.04$~meV~\cite{wipf1981neutron}, and low-temperature heat-capacity measurements gave the same $0.19\pm0.02$~meV scale~\cite{wipf1984h}. Abogoda \textit{et al.} recently revisited this O-trapped H defect with DFT NEB calculations, reconstructed the one-dimensional potential with harmonic endpoint tails, and solved the corresponding Schr\"odinger equation~\cite{Abogoda2025}. We therefore calculated the OTP for this system starting from the computed NEB path and optimizing the path to minimize the action. We estimated the harmonic tails from the endpoint force constants. Our reconstructed one-dimensional potential, shown in Fig.~\ref{fig:tunneling_validation_benchmarks}(a), gives a splitting of $0.21$~meV, consistent with experiment without any fitting of the mass. This agreement supports the OTP tunneling path approach and one-dimensional reconstruction used here to calibrate H tunneling.

To determine tunneling splittings in our ensemble approach for the amorphous oxide cells, we map each NEB tunneling profile onto an effective one-dimensional quartic double well with barrier height $V_0$ and minimum-to-minimum separation $r=2a$, retain the endpoint asymmetry separately through $\varepsilon$, and evaluate the corresponding semiclassical splitting~\cite{Garg2000},
\begin{equation}
\Delta
=4\sqrt{3}\hbar\omega
\Big(\frac{S_{0}}{2\pi\hbar}\Big)^{1/2}
e^{-S_{0}/\hbar},
\label{eq:delta_quartic}
\end{equation}
with $\omega=(8V_0/m_{\mathrm{eff}}a^2)^{1/2}$ and
$S_0/\hbar=16V_0/(3\hbar\omega)$. We use
$m_{\mathrm{eff}}=1.56\,m_{\mathrm H}$, chosen so that the reduced
NEB--WKB model for O--H/Nb reproduces the $0.21$~meV splitting obtained
from our OTP-based one-dimensional Schr\"odinger calculation from above, which involves no fitted effective mass and agrees with the
experimental splitting. This closely follows
the calibration strategy of Abogoda \textit{et al.}~\cite{Abogoda2025},
who instead calibrated their NEB-based model directly to the  experimental splitting.

As a complementary consistency check, we estimate the H-projected
tunneling mass directly from the OTP path using
$M_{\mathrm{eff}}(w)=[\Delta s(w)/\Delta x_{\mathrm H}(w)]^2$, where
$s$ is the mass-weighted coordinate and $x_{\mathrm H}$ is the
H-projected coordinate. This gives $1.08\,m_{\mathrm H}$ locally at the barrier top and approaches $\sim1.7\,m_{\mathrm H}$ when averaged over larger symmetric windows that include additional host motion, further supporting the $1.56\,m_{\mathrm H}$ choice. Figure~\ref{fig:tunneling_validation_benchmarks}(b) compares the NEB-WKB estimates with full one-dimensional Schr\"odinger calculations based on OTPs for a representative subset of Nb$_2$O$_{5-\delta}$ and Ta$_2$O$_{5-\delta}$ paths, showing promising alignment and providing an independent check of the reduced model we employ for ensemble-scale statistics.

\noindent\emph{Step 4: Statistical Analysis.---} Once tunnel splittings have been assigned to the candidate double wells, we convert the ensemble into a statistical density of microwave-active TLS. Each candidate pair is mapped to the TLS parameters $\varepsilon$ and $\Delta$ in the effective Hamiltonian~\cite{Anderson1972,Phillips1972,Mueller2019}
\begin{equation}
\hat H_{\mathrm{TLS}}=\tfrac12\varepsilon\sigma_z+\tfrac12\Delta\sigma_x,\qquad
E_{01}=\sqrt{\Delta^2+\varepsilon^2}=h f,
\label{eq:WKB}
\end{equation}
where $\varepsilon$ is the asymmetry and $\Delta$ is the tunnel splitting. The microwave-active TLS fraction is then obtained by combining the distribution of tunnel splittings with the distribution of energy asymmetries and integrating over the frequency window of interest. For the ensemble, we approximate the joint distribution as
\begin{equation}
P(\Delta,\varepsilon)
\approx
p_\Delta\!\left(\Delta\,\middle|\,|\varepsilon|<\varepsilon_{\rm cut}\right)
p_\varepsilon(\varepsilon),
\label{eq:joint}
\end{equation}
assuming that the distribution of tunnel splittings $\Delta$ is only weakly correlated with $\varepsilon$ within the low-asymmetry regime relevant to microwave-active TLS.
Integrating Eq.~\eqref{eq:joint} over the region of $(\Delta,\varepsilon)$ satisfying Eq.~\eqref{eq:WKB} within a target band $\mathcal{B}=[f_{\min},f_{\max}]$ gives $P_{\mathcal{B}}$, the fraction of valid TLSs that is microwave active within the target band. Next, to get the two-level-system density of states, we must determine the total density of TLSs. For $N_{\mathrm{min}}$ minima and $N_{\mathrm{pair}}$ unique, valid TLS pairs within
$r_{\mathrm{cut}}$ in volume $V$, the site density is $n_{\mathrm{site}}=N_{\mathrm{min}}/V$
and the average number of neighbors is $z_{\mathrm{eff}}=2N_{\mathrm{pair}}/N_{\mathrm{min}}$. Multiplying
the unique-pair density $n_{\mathrm{site}}z_{\mathrm{eff}}/2$ by the probability that a given pair is occupied by exactly one H,
$2x_{\mathrm{site}}(1-x_{\mathrm{site}})$, gives the two-level-system density of states
\begin{equation}
P_0
=\frac{
n_{\mathrm{site}}z_{\mathrm{eff}}
x_{\mathrm{site}}(1-x_{\mathrm{site}})f_{\mathrm{valid}}
P_{\mathcal{B}}
}{hf_{\max}-hf_{\min}},
\label{eq:rho_eff}
\end{equation}
where $h$ is Planck's constant.
Given this density of states, the standard low-field loss
is~\cite{martinis2005decoherence}
\begin{equation}
\tan\delta_0=\frac{\pi P_0p^2}{3\varepsilon_r\varepsilon_0},
\label{eq:loss}
\end{equation}
where $p$ is the dipole moment and $\varepsilon_r$ the relative
permittivity.

\begin{figure}[!htbp]
  \centering
  \includegraphics[width=0.9\columnwidth]{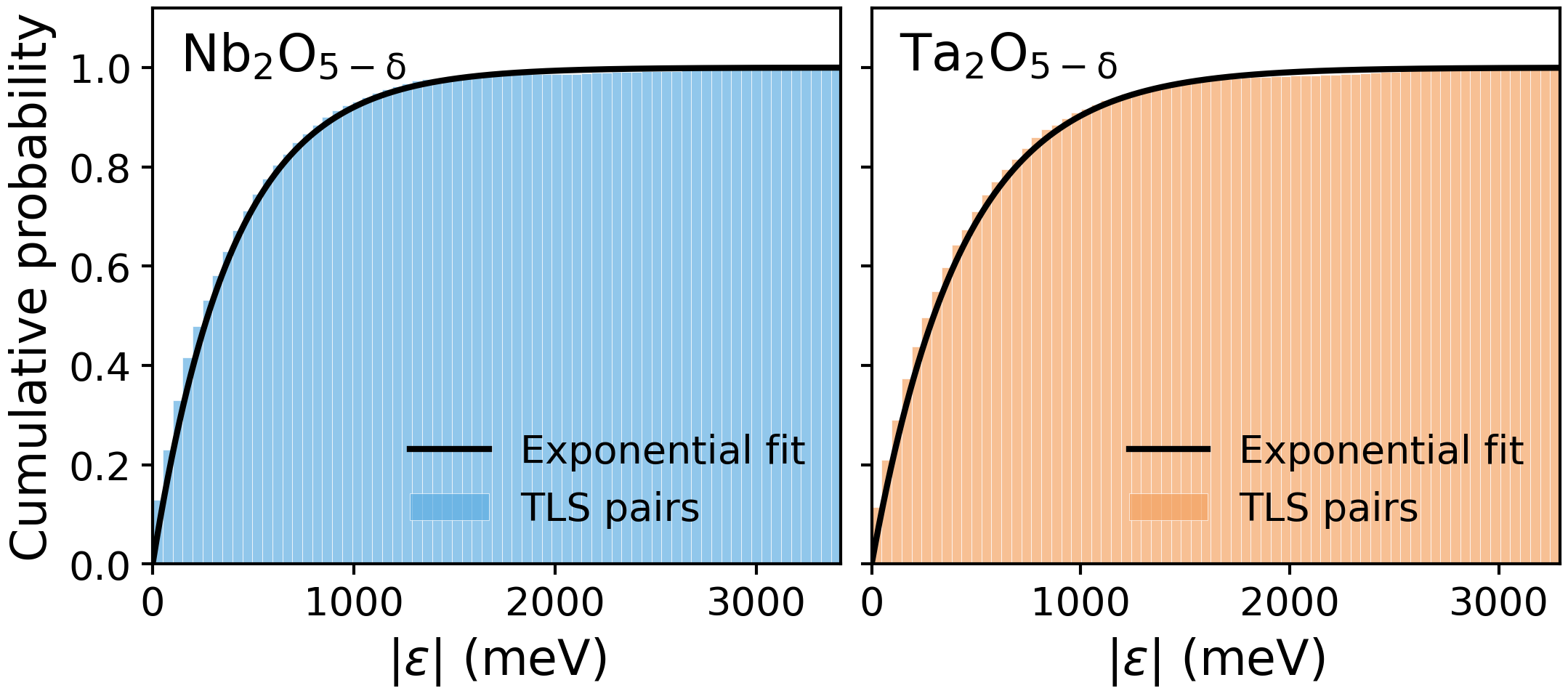}
  \caption{Hydrogen-pair asymmetry $|\varepsilon|$ distributions in amorphous Nb$_2$O$_{5-\delta}$ and Ta$_2$O$_{5-\delta}$ for all distinct-minima pairs within $r_{\mathrm{cut}}=1.65$~\AA\ (4235 Nb and 4855 Ta pairs), prior to the asymmetry cut used to select candidate tunneling pairs. Black curves show exponential fits used to extrapolate the low-$\varepsilon$ tail entering $p_\varepsilon(\varepsilon)$.}
  \label{fig:h_pair_asymmetry_distributions}
\end{figure}

For the amorphous H-pair ensembles, we take $p_\Delta(\Delta)$ in Eq.~\eqref{eq:joint} from the empirical distribution of tunnel splittings computed for the near-degenerate candidate pairs ($|\varepsilon|\le\varepsilon_{\mathrm{cut}}$). The asymmetry factor $p_\varepsilon(\varepsilon)$, by contrast, is estimated from the much larger ensemble of all distinct-minima pairs within $r_{\mathrm{cut}}$, \emph{without} the asymmetry cut; Fig.~\ref{fig:h_pair_asymmetry_distributions} in the End Matter shows these distributions, which we fit with an exponential form and extrapolate into the microwave window. This empirical $p_\Delta(\Delta)$, together with the fitted $p_\varepsilon(\varepsilon)$, determines $P_{0.1-10}$, the $0.1$--$10$~GHz value of $P_{\mathcal{B}}$ in Eq.~\eqref{eq:rho_eff}. The remaining density and loss parameters are summarized in the End Matter.

We estimate the TLS-relevant change in dipole moment between endpoint minima by calculating DFT Bader charges~\cite{henkelman2006fast} for a representative set of 36 randomly selected and rerelaxed TLS candidates. We obtain remarkably consistent mean values with narrow rms spreads: $|\Delta \mathbf{p}| = 1.11 \pm 0.12$ and $1.11 \pm 0.16$~D for Nb and Ta oxides, respectively, consistent with values inferred for TLSs in amorphous dielectrics and junction barriers. For example, Lisenfeld \emph{et al.} report 0.96--1.92 D~\cite{lisenfeld2019electric}. Taking $p \equiv |\Delta \mathbf{p}| \approx 1.11$~D, we evaluate Eq.~\eqref{eq:loss} using representative relative permittivities for the amorphous oxides, $\varepsilon_r=45$ for Nb$_2$O$_5$~\cite{Fuschillo1975,ganesan2026two} and $\varepsilon_r=21$ for Ta$_2$O$_5$~\cite{chen1997study,sethi2007structure}. The only remaining external parameter is the H concentration, which enters through $x_{\mathrm{site}}$. Our analysis therefore allows us to map our statistical ensemble directly onto predictions for loss tangents as a function of hydrogen concentration.

\begin{figure}[!htbp]
  \centering
  \includegraphics[width=1.0\columnwidth]{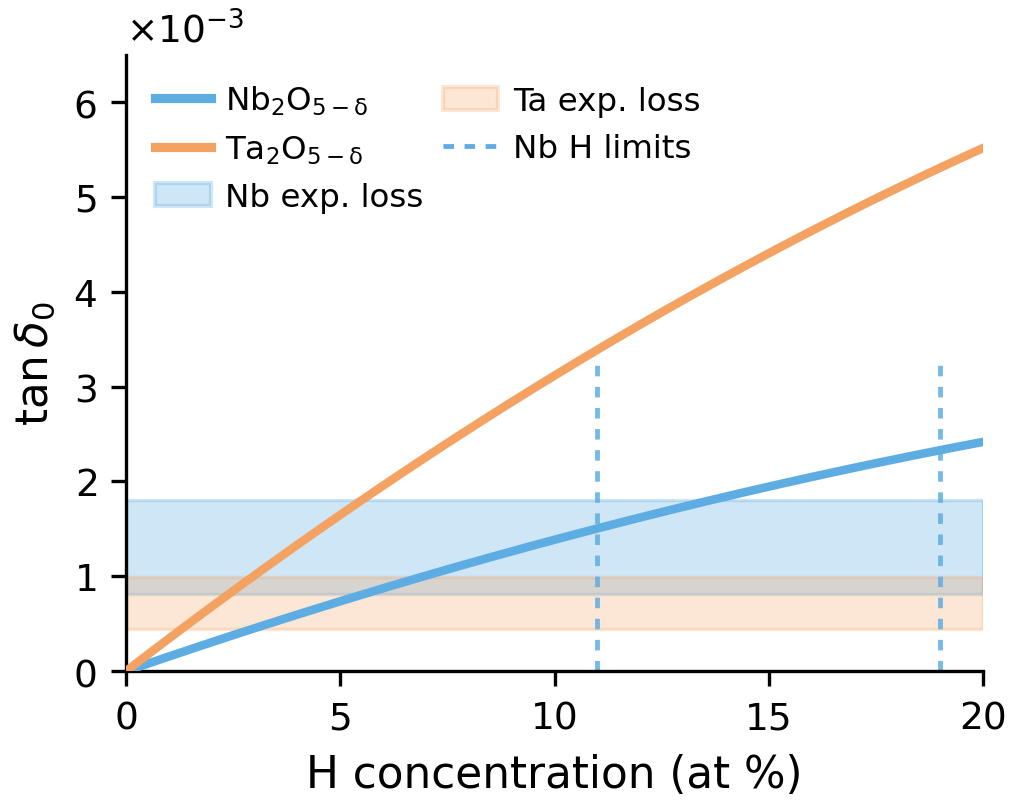}
  \caption
{Predicted low-field TLS loss tangent $\tan\delta_0$ versus hydrogen concentration
in amorphous Nb$_2$O$_{5-\delta}$ and Ta$_2$O$_{5-\delta}$.
The shaded bands show the experimentally reported Nb$_2$O$_5$ range
\cite{ganesan2026two} and the corresponding Ta$_2$O$_5$ range obtained
using the Ta/Nb loss-tangent ratio of $0.55$ reported by Goronzy
\emph{et al.}~\cite{goronzy2026comparison}. The vertical dashed lines
mark the $11$--$19$ at.\% H fractions inferred from measured elastic recoil detection (ERD)
depth profiles of Nb oxides~\cite{romanenko2011elastic}.}
  \label{fig:tls_loss_vs_h_concentration}
\end{figure}

\noindent \emph{Results and Discussion.---}
We now evaluate the H-tunneling model against the two quantitative experimental benchmarks motivating our Nb/Ta comparison: the absolute loss of Nb pentoxide and the relative microwave-active TLS spectral weight of the two pentoxides. The curves in Fig.~\ref{fig:tls_loss_vs_h_concentration} show that Nb$_2$O$_{5-\delta}$ reaches the reported range for loss tangents $\tan\delta_0$ of $8.1\times10^{-4}$ to $1.8\times10^{-3}$~\cite{ganesan2026two} for H concentrations of $5.5$--$13.6$ at.\%. This range overlaps the 11--19~at.\% concentrations observed in surface Nb$_2$O$_5$ layers of unrinsed heat-treated Nb samples~\cite{romanenko2011elastic}. Thus, the model can explain the magnitude of Nb-pentoxide loss at experimentally observed local H concentrations.

The second benchmark is the relative microwave loss of the two pentoxides. Goronzy \emph{et al.} report a
Ta$_2$O$_5$/Nb$_2$O$_5$ loss-tangent ratio of $0.55$
\cite{goronzy2026comparison}. Scaling the experimental Nb$_2$O$_5$
range by this ratio gives the Ta$_2$O$_5$ range shown in
Fig.~\ref{fig:tls_loss_vs_h_concentration}. Our model reproduces this
Nb/Ta loss contrast for Ta H concentrations of approximately
$1.3$--$2.9$~at.\%.

This inferred weaker H incorporation aligns with both our systematically higher hydrogen formation energies
in Ta oxide and H-sensitive measurements on directly comparable
surface-oxide stacks: H$^-$ ToF-SIMS counts are reported to be approximately an
order of magnitude lower in Ta-capped Nb films than in uncapped Nb
films~\cite{oh2024structure}.

The agreement with the absolute Nb loss,
the Nb/Ta loss contrast, and the observed H-uptake
trend does not uniquely identify hydrogen as the source of the TLS loss, but establishes interstitial-H
tunneling as a quantitatively viable mechanism and predicts that
suppressing hydrogen incorporation should reduce TLS loss.

\noindent \emph{Conclusions.---}
We introduced a first-principles-guided framework for predicting statistical TLS distributions from mobile impurities in disordered materials by combining MLIP-accelerated landscape sampling, targeted DFT validation, minimum-action-path and Schr\"odinger benchmarks, and statistical coarse-graining into TLS densities and dielectric loss. Applied to amorphous Nb$_2$O$_{5-\delta}$ and Ta$_2$O$_{5-\delta}$, initial screening identifies H as far more likely than oxygen or nitrogen to produce microwave-frequency tunneling. The resulting H-tunneling ensemble reproduces the measured Nb-pentoxide loss at experimentally inferred local H concentrations and the relative Nb/Ta TLS spectral weight when weaker H incorporation in Ta is taken into account. Together, these results establish interstitial-H tunneling as a quantitatively viable, though not uniquely identified, source of microwave loss and predict that suppressing hydrogen incorporation should reduce TLS dissipation. More broadly, the framework provides a route for testing other candidate impurities and motivates direct experiments relating hydrogen incorporation and removal to microwave loss.

\noindent \emph{Acknowledgments.---}
We thank Nathan Sitaraman for helpful feedback and discussions. This work was supported by the U.S. National Science Foundation through the Center for Bright Beams under award PHY-1549132 and by the Air Force Office of Scientific Research under award number FA9550-23-1-0706. Any opinions, findings, and conclusions or recommendations expressed in this material are those of the author(s) and do not necessarily reflect the views of the United States Air Force.

\bibliographystyle{apsrev4-2}
\bibliography{PRL}

\clearpage

\onecolumngrid
\section*{End Matter}
\small
\twocolumngrid

\noindent \emph{Density and loss parameters.---}
Table~\ref{tab:workflow_params} collects the material-specific quantities used in Eqs.~\eqref{eq:rho_eff} and~\eqref{eq:loss} to convert the sampled H double-well ensembles into band-averaged TLS densities and intrinsic loss tangents. The structural quantities are obtained from the amorphous-cell ensembles and H-site clustering, while $P_{0.1-10}$ comes from the empirical tunnel-splitting distribution combined with the fitted low-asymmetry tail described in the main text.

\begin{table}[!htbp]
\caption{Parameters used to convert the sampled H double-well ensembles into the band-averaged TLS density and loss estimates.}
\label{tab:workflow_params}
\centering
\scriptsize
\setlength{\tabcolsep}{3pt}
\begin{ruledtabular}
\begin{tabular*}{\columnwidth}{@{\extracolsep{\fill}}p{0.52\columnwidth}cc}
Parameter & Nb$_2$O$_{5-\delta}$ & Ta$_2$O$_{5-\delta}$ \\
\hline
Amorphous cells & 49 & 49 \\
Oxygen-deficiency range $\delta$ & 0--0.250 & 0--0.250 \\
$n_{\mathrm{host}}$ (\AA$^{-3}$) & 0.0670 & 0.0681 \\
$n_{\mathrm{site}}$ (\AA$^{-3}$) & 0.0591 & 0.0659 \\
$x_{\mathrm{site}}$ at 1 at.\% H & 0.0113 & 0.0103 \\
$z_{\mathrm{eff}}$ (pairs/site) & 1.19 & 1.24 \\
Distinct H minima & 7137 & 7825 \\
Pairs within $r_{\mathrm{cut}}$ (for $p_\varepsilon$ fit) & 4235 & 4855 \\
Candidate pairs ($|\varepsilon|\le\varepsilon_{\mathrm{cut}}$, NEB) & 123 & 150 \\
Valid double wells & 59 & 81 \\
$f_{\mathrm{valid}}$ from NEB sample & 0.480 & 0.540 \\
$P_{0.1-10}$ & $7.60\times10^{-5}$ & $6.59\times10^{-5}$ \\
$\varepsilon_r$ & 45 & 21 \\
$p$ (D) & 1.1 & 1.1 \\
\end{tabular*}
\end{ruledtabular}
\end{table}

\noindent \emph{Computational details.---}
To implement Step 1 for interstitial H, we place H on a uniform three-dimensional grid with 1~\AA\ spacing in each host structure, relax every seed to the nearest minimum, and cluster the relaxed configurations into distinct sites. We define the neutral H interstitial formation energy as $E_f(H_i)=E(\mathrm{host}+H)-E(\mathrm{host})-\tfrac12 E(\mathrm{H}_2)$. All large-scale relaxations and NEB calculations use the \texttt{fairchem} MLIP~\cite{wood2025uma,fairchem_github}, with the \texttt{uma-m-1p1}-\texttt{omat} universal model for the inorganic host materials, while representative interstitial endpoints and barrier profiles are validated with DFT using the open-source plane-wave code JDFTx~\cite{payne1992iterative,sundararaman2017jdftx}, ultrasoft pseudopotentials~\cite{garrity2014pseudopotentials}, plane-wave cutoffs of 20~Hartree for the electronic wave functions and 100~Hartree for the electron density, the PBE exchange--correlation functional~\cite{perdew1996pbe}, and Monkhorst--Pack $2\times2\times2$ $k$-point meshes~\cite{monkhorst1976}. For the bulk benchmark, we use a $4\times4\times4$ bcc Nb supercell (64 atoms). For the amorphous oxides, we use 49 distinct Nb$_2$O$_{5-\delta}$ cells and 49 distinct Ta$_2$O$_{5-\delta}$ cells spanning seven oxygen-deficiency compositions and total sizes of 162--168 atoms, taken from Pritchard \textit{et al.}~\cite{pritchard2025suppressed}.

For the O--H/Nb Schr\"odinger benchmark in Fig.~\ref{fig:tunneling_validation_benchmarks}(a), we used the NEB-seeded OTP segment as the central barrier and extrapolated the two outer regions with harmonic potentials determined from endpoint force constants, following the potential-reconstruction strategy of Ref.~\cite{Abogoda2025}. The stationary one-dimensional Schr\"odinger equation was then solved on the reconstructed potential in the mass-weighted coordinate. The oxide comparisons in Fig.~\ref{fig:tunneling_validation_benchmarks}(b) used the same potential-reconstruction procedure for selected Nb$_2$O$_{5-\delta}$ and Ta$_2$O$_{5-\delta}$ paths, providing a direct check of the NEB-WKB reduction used for ensemble-scale screening.

\begin{figure}[!htbp]
  \centering
  \includegraphics[width=0.9\columnwidth]{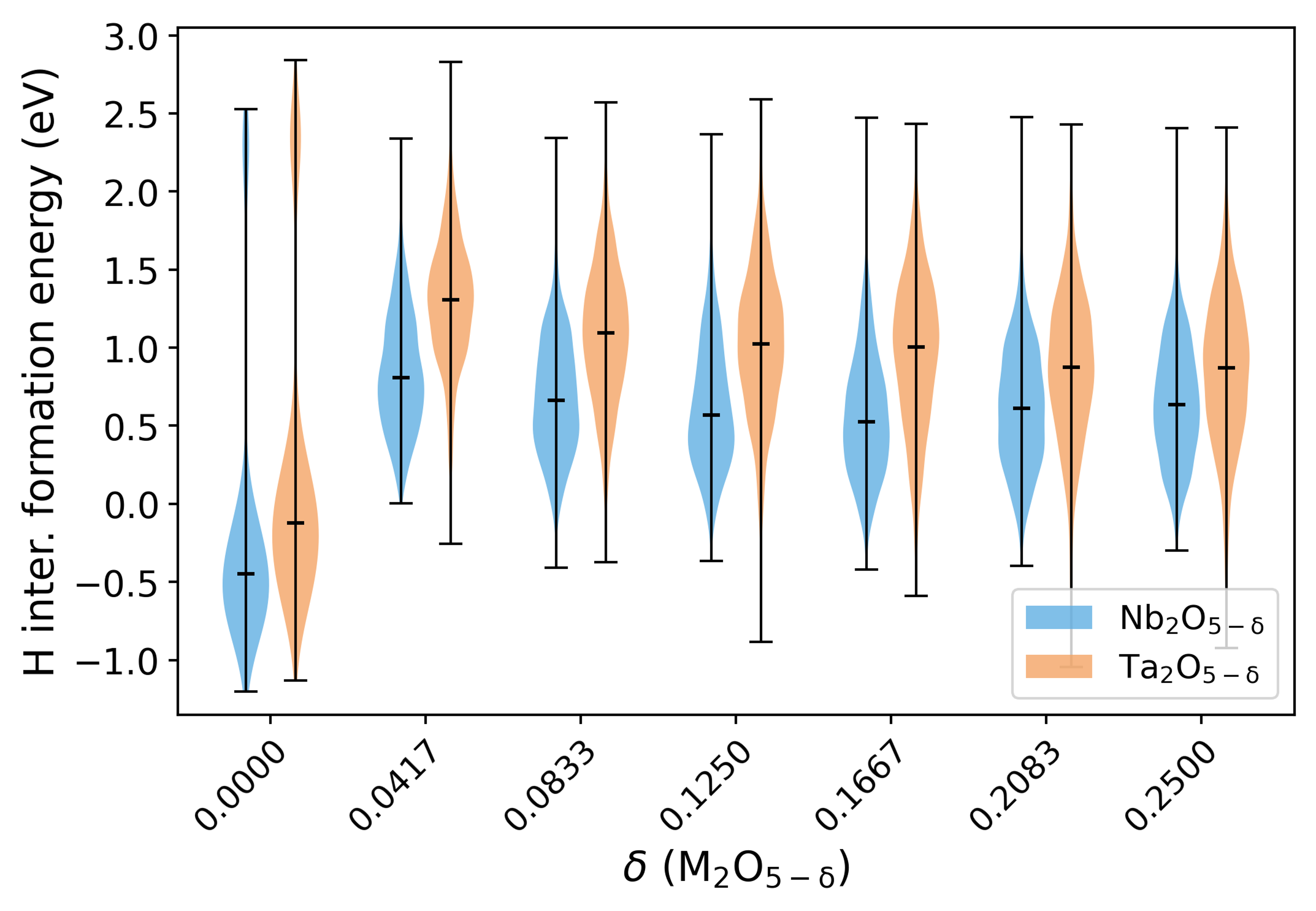}
  \caption{Distribution of relaxed H interstitial formation energies (relative to H$_2$) over the distinct sites identified in each amorphous cell as a function of oxygen-deficiency composition $\delta$. Black markers show the median and interquartile range.}

  \label{fig:h_interstitial_formation_energies}
\end{figure}

For completeness, Fig.~\ref{fig:h_interstitial_formation_energies} summarizes the full distribution of relaxed H interstitial formation energies as a function of oxygen deficiency in amorphous Nb$_2$O$_{5-\delta}$ and Ta$_2$O$_{5-\delta}$. The black markers denote the median and interquartile range over the distinct relaxed H sites identified in each structural ensemble. The distributions show that H formation energies in Nb$_2$O$_{5-\delta}$ remain systematically lower than in Ta$_2$O$_{5-\delta}$ across the full $\delta$ range. Within the H-tunneling interpretation, this thermodynamic trend supports lower H incorporation as a contributor to lower Ta loss and is consistent with experimentally inferred stronger H uptake in Nb-based native oxides.

\end{document}